\documentclass[twocolumn,preprintnumbers,amsmath,amssymb]{revtex4}

\usepackage{graphicx}
\usepackage{dcolumn}
\usepackage{bm}

\begin{document}

\preprint{APS/123-QED}

\title{Photoluminescence measurements in Be-$\delta$-doped back-gate induced quantum well}

\author{M. Yamaguchi$^{1,3}$}
\email{m-yama@nttbrl.jp}
\author{S. Nomura$^{2,3}$}
\author{D. Sato$^{2,3}$}
\author{T. Akazaki$^{1,3}$}
\author{H. Tamura$^{1,3}$}
\author{H. Takayanagi$^{1,3}$}

\affiliation{$^{1}$NTT Basic Research Laboratories, NTT Corporation, 3-1 Morinosato-Wakamiya, Atsugi-shi, Kanagawa 243-0198, Japan\\
$^{2}$Institute of Physics, University of Tsukuba, 1-1-1 Tennodai Tsukuba, Ibaraki 305-8571, Japan\\
$^3$CREST, Japan Science and Technology Agency, Honmachi, Kawaguchi, 332-0012, Saitama, Japan}%
\homepage{http://www.brl.ntt.co.jp/spintronics/}

\date{\today}

\begin{abstract}
The photoluminescence (PL) spectra of a two-dimensional electron system induced in a Be-$\delta$-doped GaAs$/$AlGaAs quantum well (QW) with a back gate are measured. The electron density is controlled from 1$\times$10$^{9}$cm$^{-2}$ to 2.5$\times$10$^{11}$cm$^{-2}$ by changing the back gate voltage. There is a linear increase in the acceptor PL spectrum around 1.49 eV with an increase in the back gate voltage and the PL disappears from the exciton bound to neutral donors (D$^{0}$X) around 1.51 eV at 1.2$\times$10$^{10}$cm$^{-2}$. 
\end{abstract}

\maketitle

\begin{figure}[t]
\begin{center}
\includegraphics[width=7cm]{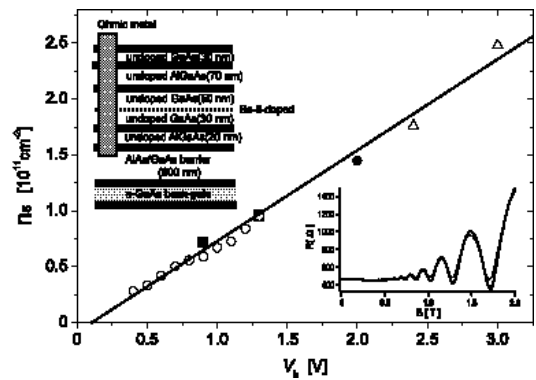}
\end{center}
\caption{\label{fig:Fig1} The electron density estimated from transport SdH measurements (open triangles), and optical SdH measurements (filled squares). The filled circle is an estimation from the Landau fan diagram. The solid line is a fit of these 5 points with the linear function provided in the text. The open circles are the electron density estimated from the line width of the PL spectrum.  The inset in the upper left corner is the layer structure and that in the lower right corner is the transport SdH oscillation at $V_b$=3.0~V. The thick and thincurves show measurements under 25 $\mu$W cm$^{-2}$ laser irradiation and for a dark condition, respectively.}
\end{figure}
 
\begin{figure}[t]
\includegraphics[width=5cm]{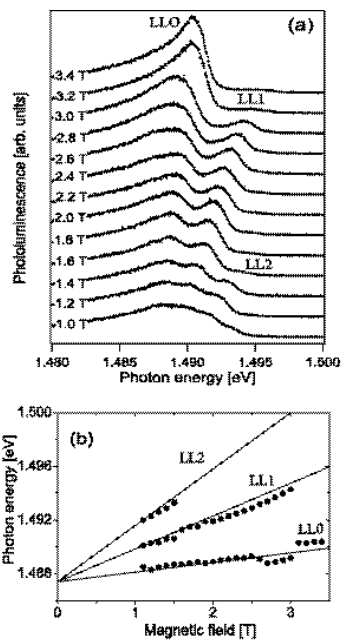}
\caption{\label{fig:Fig2} (a) The magnetic field dependence of the PL spectrum at $V_b$=2.0~V. The three Landau peaks, LL0, LL1, and LL2 are visible abeve 1.2 T. (b) The Landau fan diagram. The solid line is the calculated Landau level spacing with $m_e$=0.067 $m_0$.}
\end{figure}

\begin{figure}[t]
\includegraphics[width=5cm]{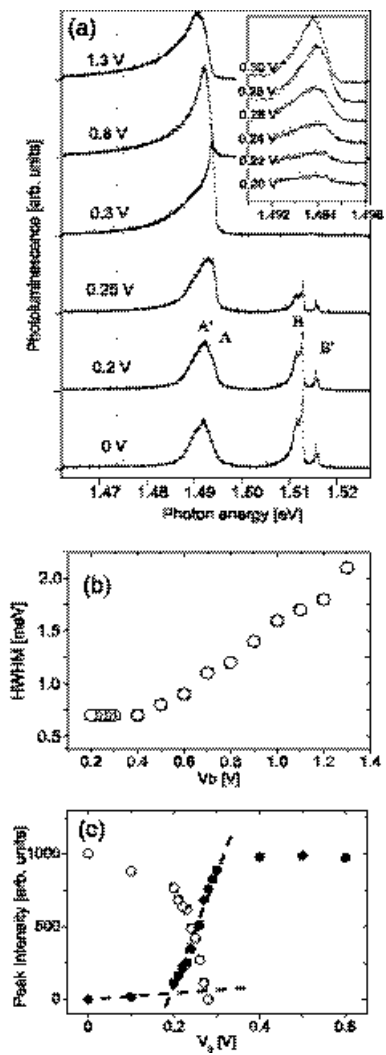}
\vskip 0.5cm
\caption{\label{fig:Fig3} (a) The PL spectra at $V_b$=0, 0.2, 0.25, 0.3, 0.8, and 1.3~V at B=0~T. Peaks A', A, B, and B' are regarded as the free-to-acceptor-impurity transition, the free-to-Be-acceptor transition, D$_0$X and X$_0$, respectively. Peak A after subtracted peak A$'$ is plotted in the inset. Peak A increases with $V_b$ around 0.2~V and peaks B and B' are not observed at $V_b$=0.3~V. (b) The back-gate dependence of the HWHM of peak A. The HWHM is constant below $V_b$=0.4~V, and increases above $V_b$=0.4~V. (c) The back-gate dependence of the height of peak A (filled circles) and peak B (open circles). The dashed lines are guides for the eye.}
\end{figure}

High controllability of the electron density and absence of scattering by ionized donors make back-gated quantum well (QW) structures advantageous for exploring novel physics in a dilute two-dimensional electron system (2DES). The 2DES induced by the back gate in an undoped GaAs$/$AlGaAs heterostructure, which shows high mobility at low electron density~\cite{Hirayama98}, is promising for investigating not only the physical properties in the low-density regime such as a metal insulator transition and Wigner crystallization, but also new device applications such as artificial magnets~\cite{ShiraishiAPL,TamuraPRB}. Transport measurements of such system have already been reported in the low electron density regime~\cite{Hirayama98,Kawaharazuka01,Lilly03}. However, there have been difficulties in measuring the transport properties when the electron density is extremely dilute, since the 2DES tends to localize and become insulating. Optical method is suitable for measuring the electrical properties of the dilute 2DES, because it can reach not only the conductive regime but also the insulating regime. 
This paper reports the photoluminescence (PL) measurement of the 2DES induced by the back gate. We focus on the PL from the recombination of the electrons with a hole bound to the negatively charged acceptor. The advantage of this method is that the optical selection rule in the recombination is relaxed due to the strong localization of a hole at the acceptor site and the PL spectrum provides a direct method for investigating the energy spectra of electrons in 2DES~\cite{Kukushkin89,Hawrylak92,Nomura03a}.

The layer structure of the sample is shown in the inset of Fig.\ref{fig:Fig1}. The layers consist of a Si-doped GaAs substrate, overgrown by molecular beam epitaxy with a 1000 nm heavily-Si-doped GaAs layer, which acts as a back gate, followed by a barrier layer comrising 150 periods of GaAs(2-nm thick)/AlAs(2-nm thick) superlattice and 20 nm of Al$_{0.3}$Ga$_{0.7}$As. Subsequently, 80 nm of GaAs is deposited as a channel, followed by 70 nm of Al$_{0.3}$Ga$_{0.7}$As, and this is capped by 30 nm of undoped GaAs. 2$\times$10$^{10}$cm$^{-2}$ Be atoms are $\delta$-doped 30 nm from the bottom of the channel as acceptors. The sample is processed into a 1 mm$\times$1 mm mesa and an AuGeNi ohmic contact is attached at the mesa edge. A back gate voltage ($V_b$) is applied between the ohmic contact and the back gate.
The PL is measured by exciting the samples at 532 nm with a continuous wave at an induced power density of 25 $\mu$Wcm$^{-2}$ at 1.7 K. The PL from the sample is dispersed through a 55-cm monochrometer and detected by a charge-coupled device cooled by liquid nitrogen.

Before optical measurements are undertaken, the electron density is estimated from the Shubnikov-de Haas (SdH) measurement for the same wafer at $V_b$=2.4~V and 3.0~V.  The SdH oscillation signal at $V_b$=3.0~V is plotted in the inset of Fig.\ref{fig:Fig1}, both under 25 $\mu$Wcm$^{-2}$ laser irradiation and in a dark condition. The electron density under irradiation is a little larger than that in the dark condition. The difference between them is less than 3\% in contrast to Si-doped single-heterostructure samples where the photo-excited carriers compensate for the charge of the ionized donors, resulting in a decrease in the electron density~\cite{Kukushkin89}. This is considered to be due to the lack of a donor layer in our sample.

Figure \ref{fig:Fig2}(a) shows the magnetic field dependence of the PL spectrum at $V_b$=2.0~V. The three peaks correspond to the Landau levels LL0, LL1, and LL2. The peak heights of the higher Landau levels LL1 and LL2 decease with magnetic field and disappear at B=1.5~T  for LL2 and 3~T for LL1. The peak energies of the Landau levels are plotted in Fig.\ref{fig:Fig2}(b). The solid lines represent the Landau level spacing calculated with $m_e$=0.067$m_0$. The jumps in the peak energies of LL1 and LL0 are observed at B=1.5~T and 3~T, respectively, where the higher Landau peaks disappear. From the Landau fan diagram in Fig.\ref{fig:Fig2}(b), the estimated electron density at $V_b$=2.0~V is 1.45 $\times 10^{11}$ cm$^{-2}$,  which is also plotted in Fig.\ref{fig:Fig1} as well as the lower electron densities estimated at $V_b$=1.3~V and 0.8~V from optical SdH measurements. The electron densities plotted in Fig.\ref{fig:Fig1} for five different $V_b$ values estimated from the transport and optical measurements are reasonably fitted by a single line as
\begin{equation}
n_s=-9.4 \times10^9+(8.2\times10^{10}) V_b \  \rm {cm}^{-2}.
\label{eq:density}
\end{equation}
        
The PL spectra at 0 T for different $V_b$'s are shown in Fig.\ref{fig:Fig3}(a). Peaks A and A$'$ are attributed to the recombination of electrons with acceptors and peaks B and B$'$ are attributed to the exciton bound to neutral donors (D$_0$X) and free exciton (X$_0$). Between $V_b$=0.2~V and 0.3~V, peaks B and B$'$ abruptly disappear and peak A grows. Peak A is much narrower than peak A$'$. Therefore, peak A is regarded as the PL of the recombination between the 2DES formed in the QW and Be-$\delta$-doped acceptors and peak A$'$ is probably due to the recombination with unintended impurities. In order to focus on the growth of peak A, the inset shows the spectra between $V_b$=0.2~V and 0.3~V after subtracting the spectum of $V_b$=0~V, since the intensity of peak A is almost 0 at $V_b$=0~V and the change in peak A$'$ between $V_b$=0.2~V and 0.3~V is smaller than that in peak A. 

The half width at half maxima (HWHM) of the high energy side of peak A are plotted in Fig.\ref{fig:Fig3}(b). The HWHM increases linearly with $V_b$ above 0.4~V, while it remains constant at 0.7 meV below 0.4~V, which reflects the distribution of the $\delta$-doped acceptor level. The increase in the linewidth is attributed to the increase in the energy difference between the band edge ($E_g$) and the Fermi level ($E_f$). 
If the FWHM (=2$\times$HWHM) is regarded as the difference between $E_f$ and $E_g$, the electron density calculated from the FWHM is almost 20\% larger than that calculated in Eq.~(\ref{eq:density}). 
If we assume that the full width of the 2/3 peak height corresponds to $E_f-E_g$, it is in good agreement with Eq.~(\ref{eq:density}) as plotted in Fig.\ref{fig:Fig1}. Since it is difficult to determine E$_g$ and E$_f$ from the spectrum because of the existence of peak A$'$, the number 2/3 itself cannot be justified by any physical reason. However, the agreement of the $V_b$ dependence of the linewidth with Eq.~(\ref{eq:density}) guarantees that Eq.~(\ref{eq:density}) is valid at least down to 2.5$\times$10$^{10}$cm$^{-2}$ at $V_b$=0.4~V. 

 The peak intensities of peaks A and B are plotted in Fig.\ref{fig:Fig3}(c). The peak intensity of peak A increases with $V_b$ from 0 to 0.4~V and the slope changes around $V_b$=0.18~V. (The two slopes below and above $V_b$=0.18~V are guided by dashed lines in Fig.\ref{fig:Fig3}(c)). Above $V_b$=0.18~V, the peak intensity increases linearly with $V_b$ and saturates around $V_b$=0.3~V, while the intensity of peak B decreases and disappears at $V_b$=0.27~V. This result indicates that $V_b$=0.18~V is the threshold voltage at which the 2DES in the QW starts to be filled by means of the back gate operation. This threshold voltage $V_b$=0.18~V is in reasonable agreement with Eq.~(\ref{eq:density}). Peak A  below V$_b$=0.18 V can be regarded as the recombination of photo-excited or thermally activated electrons in the QW. The electron density at $V_b$=0.18~V is estimated to be 1$\times$ 10$^9$ cm$^{-2}$ according to the ratio between the intensities of  peak A at $V_b$=0.18~V and $V_b$=0.3~V.  When the 2DES is induced in the QW by the back gate, the exciton peaks B and B$'$ decrease because the binding force between an electron and a hole is screened by the 2DES and the acceptor PL becomes dominant. With increasing $V_b$, peak A saturates around $V_b$=0.3~V. The electron density of 1.5$\times$ 10$^{10}$ cm$^{-2}$ at $V_b$=0.3~V is comparable to the density of the $\delta$-doped Be (2$\times$ 10$^{10}$ cm$^{-2}$). The $V_b$ dependence of the PL intensity is dominated by the population of the optically generated hole above $V_b$=0.3~V.

In summary, we observed the PL spectrum from a Be-$\delta$-doped gate induced QW depending on the electron density. We distinguished the density of the 2DES induced by the back gate operation about 1$\times$ 10$^{9}$ cm$^{-2}$. In addition, the measurement confirms that a uniform 2DES can be induced in the back-gated hetero-junction as large as 1 mm$^2$. Our results show that a Be-$\delta$-doped QW with a back gate is suitable for studying a dilute 2DES by the optical method.

We thank J. Nitta and Y. Harada for useful discussions.

\end{document}